\documentclass[letterpaper,12pt]{article}
\pdfoutput=1

\usepackage{jheppub}
\usepackage{amsmath}
\usepackage{graphicx}
\usepackage{subfig}
\usepackage{cancel}
\usepackage[dvipsnames]{xcolor}
\usepackage{color}
\usepackage{mathrsfs}

\usepackage{amssymb}
\usepackage{epstopdf}
\newcommand{\WdW}{W\!dW}

\newpage

\title{Holographic complexity is nonlocal}

\author[a]{Zicao Fu,}
\author[b]{Alexander Maloney,}
\author[a]{Donald Marolf,}
\author[b]{Henry Maxfield,}
\author[c]{Zhencheng Wang}

\affiliation[a]{Department of Physics, University of California, Santa Barbara, CA 93106, USA}
\affiliation[b]{Department of Physics, McGill University, 3600 rue Universit\'e, Monteal H3A 2T8, Canada}
\affiliation[c]{School of Physics, Nankai University, Tianjin 300071, China}

\emailAdd{zicaofu@physics.ucsb.edu}
\emailAdd{maloney@physics.mcgill.ca}
\emailAdd{marolf@physics.ucsb.edu}
\emailAdd{henry.maxfield@physics.mcgill.ca}
\emailAdd{wangzhencheng@mail.nankai.edu.cn}

\abstract{
We study the ``complexity equals volume'' (CV) and ``complexity equals action'' (CA) conjectures by examining moments of  of time symmetry for $\rm AdS_3$ wormholes having $n$ asymptotic regions and arbitrary (orientable) internal topology.   For either prescription, the complexity relative to $n$ copies of the $M=0$ BTZ black hole takes the form $\Delta C = \alpha c \chi $, where $c$ is the central charge and $\chi$ is the Euler character of the bulk time-symmetric surface. The coefficients $\alpha_V = -4\pi/3$, $\alpha_A = 1/6 $  defined by CV and CA are independent of both temperature and any moduli controlling the geometry inside the black hole. Comparing with the known structure of dual CFT states in the hot wormhole limit, the temperature and moduli independence of $\alpha_V$, $\alpha_A$  implies that any CFT gate set defining either complexity cannot be local. In particular, the complexity of an efficient quantum circuit building local thermofield-double-like entanglement of thermal-sized patches does not depend on the separation of the patches so entangled. We also comment on implications of the (positive) sign found for $\alpha_A$, which requires the associated complexity to decrease when handles are added to our wormhole.
}

\begin{document}
\maketitle

\section{Introduction}
\label{sec:Introduction}

The AdS/CFT correspondence has proven to be one of the most powerful tools in the study of quantum gravity, and has  recently led to a remarkable set of relationships between quantum information theory and quantum gravity.
These relationships have their roots in the Bekenstein-Hawking formula -- $S={A/4 G}$ -- which relates an information theoretic quantity, namely entropy, to the geometry of a black hole in the form of the horizon area.  Entropy, however, is only a coarse probe of a quantum state.  It is natural to ask whether other more refined information theoretic notions are encoded in other geometric features of a black hole.
Some of the most interesting recent proposals relate the \emph{computational complexity} of a state to features of the geometry behind the event horizon of a black hole  \cite{Susskind:2014rva,Stanford:2014jda,Roberts:2014isa,Brown:2015bva,Brown:2015lvg}.  In this paper we will consider these proposals in a simple theory of gravity, in three dimensions with negative cosmological constant, which will allow us to compare these different proposals in a context where it is possible to make precise statements about a wide class of gravitational states with known CFT duals.

The computational complexity $C$ of a quantum state is, at least intuitively, easy to define.  It is simply the number of elementary unitary operations required to construct a desired state from a given reference state.  Even for simple quantum systems, such as finite spin chains, this definition is still ambiguous as it depends on a choice of reference state and of which unitary operators one chooses to regard as ``elementary'', as well as a notion of the accuracy required for the prepared state.  Nevertheless, many features of computational complexity, such as the growth of complexity with time, are expected to be independent of these choices, possibly up to some overall normalization of $C$.

In the context of holography there are two competing conjectures which relate this complexity to features of the bulk geometry.
Both of these conjectures depend on the choice of space-like slice in the boundary on which the quantum state is defined, as well as on the geometry of the bulk dual. The ``complexity equals volume'' (CV) conjecture \cite{Stanford:2014jda,Roberts:2014isa} is that
\begin{equation}
C = {V \over G \ell}~.
\end{equation}
Here $V$ is the volume of the maximum volume space-like slice in the bulk which is anchored on our boundary slice.  One feature  of this proposal is that it requires an additional choice of a length scale $\ell$, which is taken to be the AdS radius.
This additional choice is avoided in the second conjecture, ``complexity equals action'' (CA) \cite{Brown:2015bva,Brown:2015lvg}, which states that
\begin{equation}
\label{eq:CA}
C = {I \over \pi \hbar}~.
\end{equation}
Here $I$ is the action of the Wheeler-DeWitt patch, which is defined as the union of all space-like slices which are anchored on our boundary slice.  The boundary of this patch is a collection of null slices in the bulk, and in general one must work out the boundary terms in this action carefully \cite{Lehner:2016vdi}; the boundary contributions can in some cases be as large as the bulk volume contribution.

In this paper, we study the consequences of these conjectures for a class of states in holographic two-dimensional conformal field theories, with bulk duals given by ``wormholes'', or ``geons''. These are locally AdS spacetimes with interesting global structure, interpreted as black holes with one or more distinct exterior regions, with a non-traversable wormhole behind the horizons, which may have nontrivial topology. Relative to an appropriate reference state dual to a set of $M=0$ BTZ black holes, we will find that the complexity depends \emph{only} on the topology of space, being proportional to its Euler character, despite the existence of many continuous moduli defining the state. We will then explore two consequences of this result. Firstly, the moduli independence, along with an understanding of the dual state, imply that the appropriate definition of complexity should be nonlocal. Secondly, the constant of proportionality of the Euler character has opposite signs for CV and CA conjectures, with the consequence for CA complexity that states with more complicated topology have lower complexity. The fact that complexity is bounded below (by zero) suggests an upper bound on the genus of space for a given energy.

It is  natural to incorporate locality into notions of complexity
for local quantum field theories, for example as discussed in section 5 of \cite{Jefferson:2017sdb}.  For instance, in a lattice model one could demand that each fundamental gate preparing the state may couple only locally, perhaps setting a maximum number of lattice sites within a set of some diameter $D$ on which it may act, or introducing penalty factors assigning higher complexity to gates coupling more distant sets of lattice sites.  It is thus of interest to ask whether this could be the case for gate sets defining notions of CFT complexity satisfying either CV or CA. This question was not addressed in \cite{Stanford:2014jda,Roberts:2014isa,Brown:2015bva,Brown:2015lvg}, but the states we consider provide some insight.

At least in the hot wormhole limit of \cite{Marolf:2015vma}, the CFT states dual to wormhole spacetimes locally resemble the thermofield double state, which describes the CFT living on two disconnected spaces, but entangled with one another in a local way. The geons generalize this to allow for similar local entanglement, but with more complicated arrangements, with some region entangled with another either in a disconnected part of space, or in some nearby or distant region in the same connected component of space. Varying certain moduli gives control over which regions of the CFT are entangled, in particular controlling their spatial separation.  Since we find that both CV and CA predict that the complexity does \emph{not} depend on this separation scale, this suggests that such complexities do not penalize gates that couple distant parts of the CFT.

We also consider what happens to the complexity as we change the spatial topology of the bulk geometry. This has a simple answer, that adding a handle will alter the complexity by a fixed number times the central charge $c$, increasing in the case of CV, and decreasing in the case of CA. This gives a sharp distinction between the two conjectures, perhaps favoring one over the other, or indicating that they measure different notions. One might immediately think it is problematic for CA that more complicated topologies have lower complexities; it is counterintuitive, and more sharply, adding sufficiently many handles could cause the complexity to become negative. However, we will argue that a direct contradiction along these lines can be avoided. We expect there to be a UV-sensitive, extensive constant in the complexity, so to cancel this requires a genus that scales with the UV-cutoff scale. But when the genus is this large, the theory at the given value of the cutoff may no longer adequately describes the physics of the geon state, evading a contradiction. This argument requires an upper bound for genus at given energy, which we will show is enforced by a Hawking-Page type phase transition in a symmetric example, is consistent with the hot limit, and which may be true more generally.

We begin in section \ref{sec:overview} with an overview of the wormhole spacetimes to be studied and their dual CFT states.  Section \ref{sec:GenusBH} then uses both CV and CA to evaluate the $t=0$ complexities of an $n$-boundary wormhole relative to that of $n$ massless BTZ black holes computed using CV and CA.  We find $\Delta C = \alpha c \chi$ for central charge $c$ and  Euler character $\chi$ of the bulk $t=0$ surface, with coefficients $\alpha_V = -4\pi/3$, $\alpha_A = 1/6$ independent of temperature and also independent of any moduli controlling the geometry in the causal shadow.  Both $\alpha_V$, $\alpha_A$ are order one constants, though they differ in sign ($\alpha_V < 0$, $\alpha_A >0$).  We then argue in section \ref{sec:Discussion} that any CFT gate set defining either complexity necessarily contains bi-local gates acting at points with arbitrary separation. In particular, the complexity of efficient quantum circuits building local thermofield-double-like entanglement of thermal-sized patches must not depend on the separation between patches.  Finally, in section \ref{genusBound} we explore the implications of the sign found for $\alpha_A$, which requires that the associated complexity decrease as handles are added to our wormhole. Positivity of complexity then requires the number of handles of the interior geometry to be bounded linearly in the temperature, $g\leq k T$ for some constant $k$ of order one, and in particular with $k$ independent of $c$. We give evidence that this bound is enforced by a phase transition to a bulk geometry of lower genus.

\section{Overview of wormhole spacetimes and dual CFT states}
\label{sec:overview}

We probe the CV and CA conjectures by studying the volume and action for AdS$_3$ wormhole spacetimes at a moment of time-symmetry.  Such spacetimes are quotients of global AdS$_3$ by discrete groups of AdS$_3$-isometries that preserve a time-reflection symmetry.   In general, the quotients have $n$ distinct asymptotic regions, each with conformal boundary $S^1 \times \mathbb{R}$, and may also feature complicated topology inside the wormhole. The domain of outer communication of any such boundary has precisely the same geometry as some static BTZ black hole of some mass $M \ge 0$.

The main features of the full Lorentz-signature spacetime may be understood by studying the ($t=0$) surface invariant under the time-reflection symmetry.  This surface is a quotient of the hyperbolic plane, having $n$ asymptotic regions each with conformal boundary $S^1$, forming a hyperbolic surface with $n$ boundaries.  The event horizon bifurcation surface associated with a given boundary is the geodesic homotopic to that boundary in the $t=0$ surface. The surface can have any topology, but the region outside the horizon near any given boundary is always topologically a cylinder. Away from $t=0$, the black hole interior evolves toward a BTZ-like singularity.  See \cite{Brill:1995jv,Aminneborg:1997pz,Brill:1998pr} for more detail.

\begin{figure}[t]
\begin{center}
\includegraphics[width=0.3\textwidth]{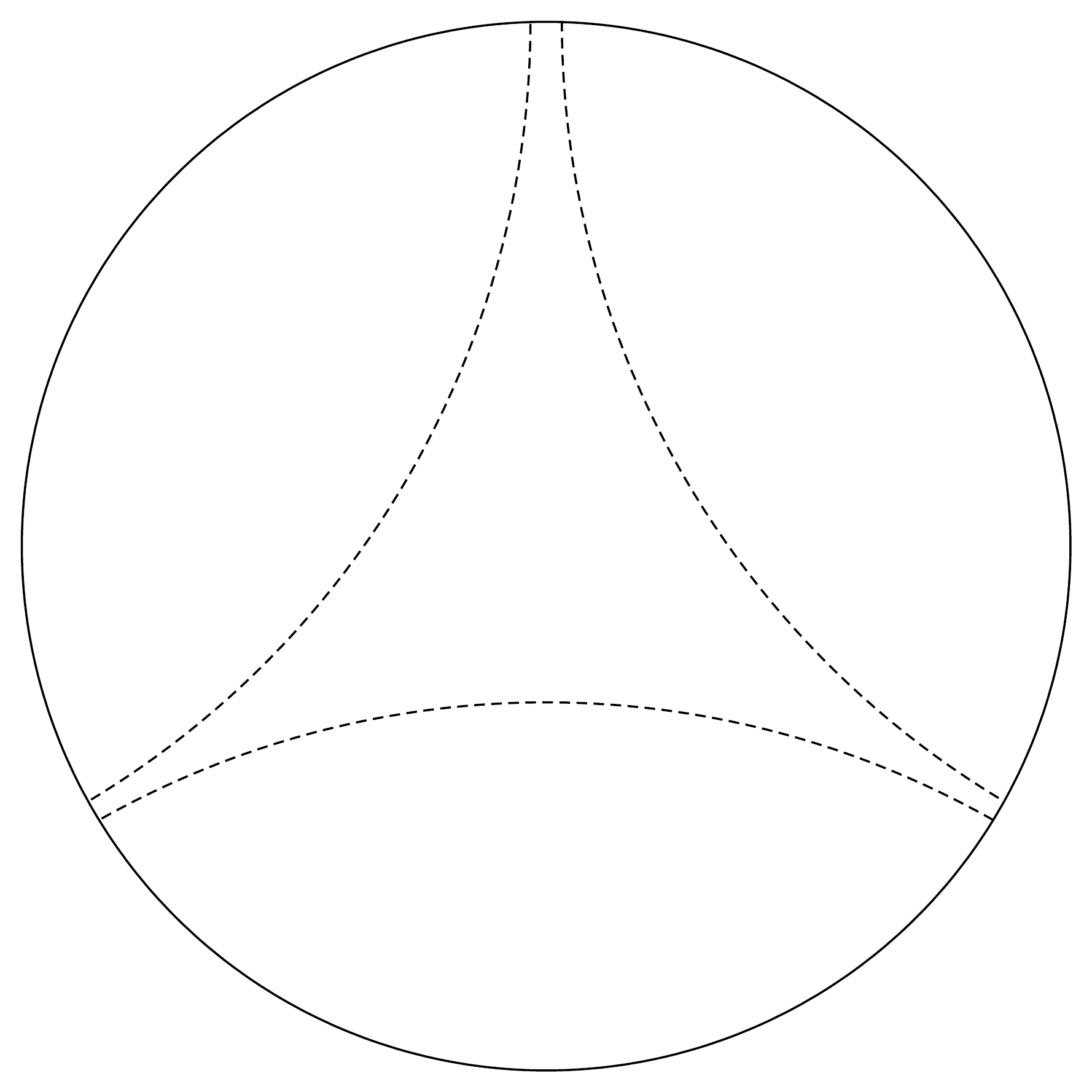}
\includegraphics[width=0.3\textwidth]{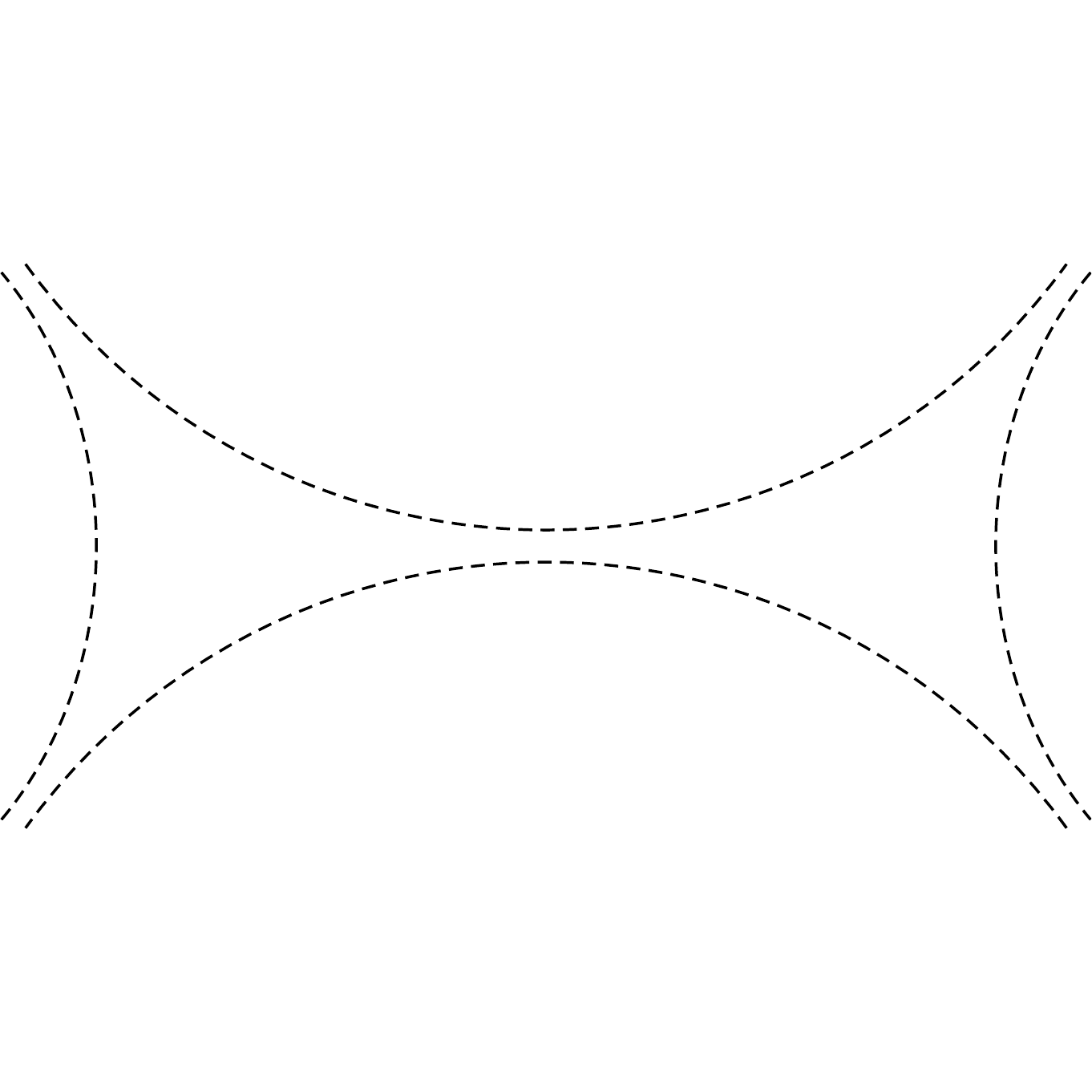}
\includegraphics[width=0.3\textwidth]{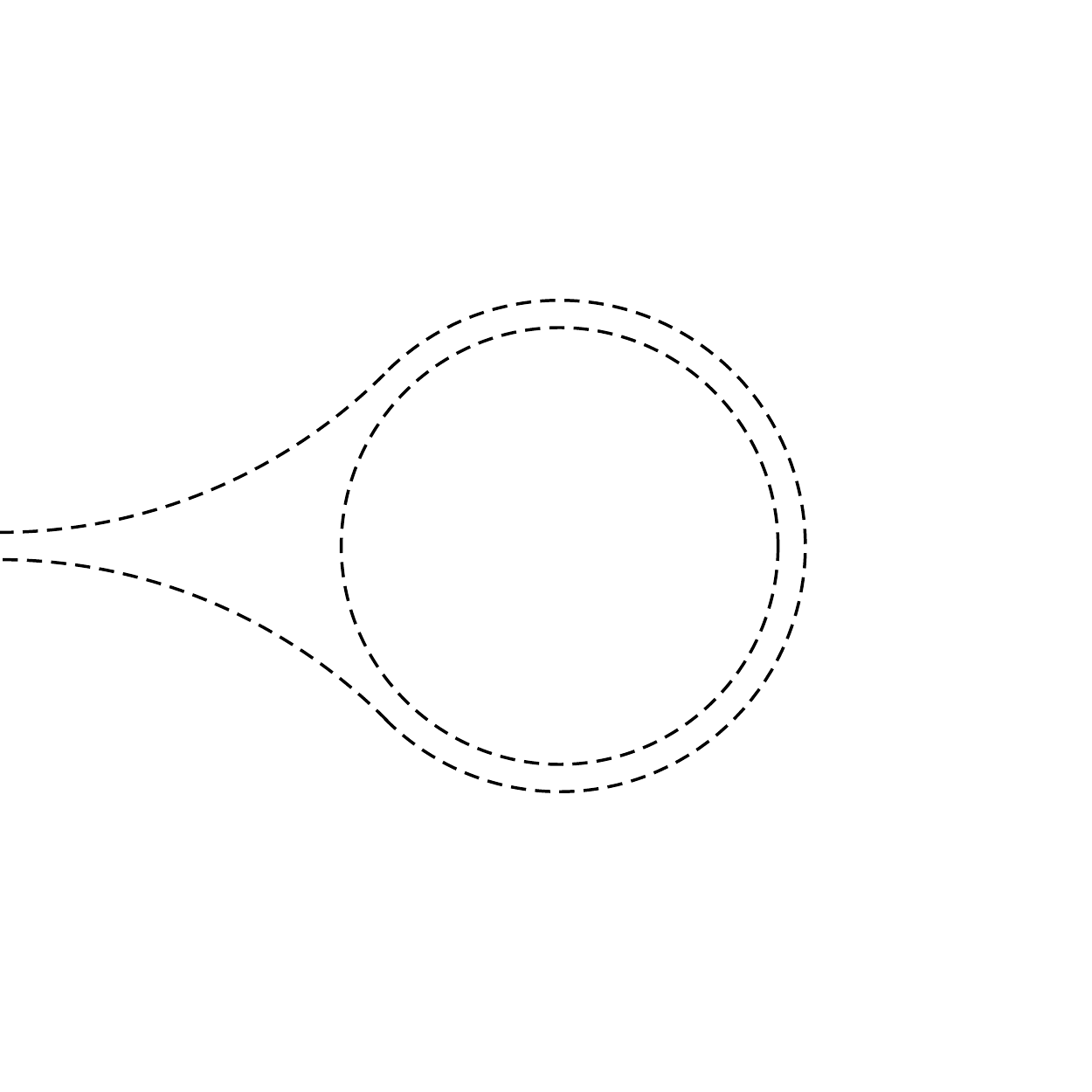}
\end{center}
\caption{{\bf Left:} The non-compact equilateral hyperbolic triangle geometry typical of the ``islands'' behind the horizon (solid lines) as a subset of the Poincar\'e disk. {\bf Center:} Rough sketch of two such islands connected by a long tendril. {\bf Right:} Rough sketch of an island with two of its own tendrils connected to each other.}
\label{fig:islands}
\end{figure}

An important feature of these spacetimes is that they are dual to known CFT states, at least in certain regions of moduli space, constructed as a path integral over a Riemann surface with boundary, as in \cite{Krasnov:2000zq, Maldacena:2001kr, Skenderis:2009ju, Balasubramanian:2014hda}.  The most explicit characterization is available in the hot wormhole limit of \cite{Marolf:2015vma}, in which the temperature of the black hole is large in AdS units and all closed geodesics within the bulk $t=0$ surface are much longer than the AdS scale $\ell$. In this limit the region behind the horizon degenerates into a finite number of ``islands'', typically\footnote{For certain topologies, or more generally on special surfaces in moduli space, several of these fundamental islands can merge into a single larger one.} with each approximating the non-compact equilateral hyperbolic triangle shown in figure \ref{fig:islands} (left) but connected along their long tendrils as shown in figure \ref{fig:islands} (center, right).

\begin{figure}[t]
\begin{center}
\includegraphics[width=0.5\textwidth]{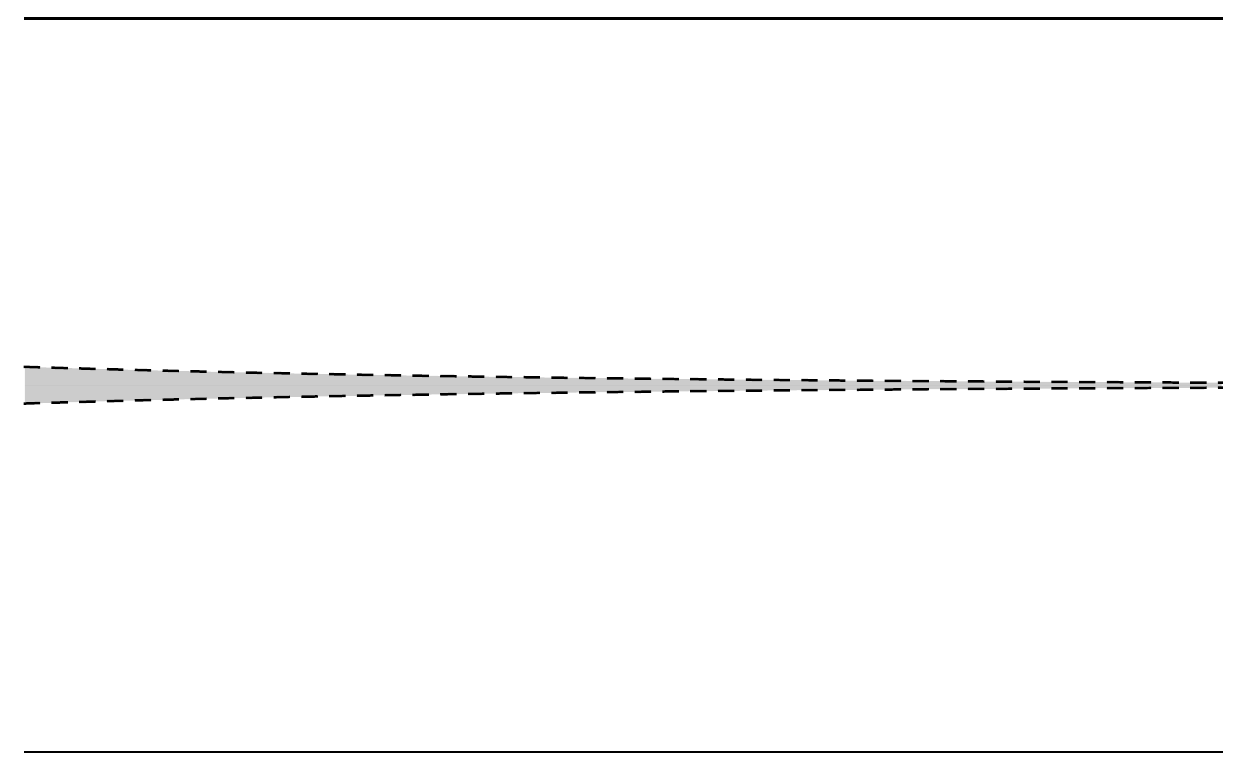}
\end{center}
\caption{In this region, the causal shadow region (shaded) behind the horizon is exponentially thin. The spacetime approximates that of the two-sided BTZ black hole, and the dual CFT state is exponentially close to the thermofield double. Solid lines are the asymptotically (locally) AdS boundaries.}
\label{fig:BTZRegion}
\end{figure}

Far out along these tendrils the causal shadow region behind the horizon becomes thin, its width decaying exponentially with the distance to the nearest island, as shown in figure \ref{fig:BTZRegion}.  The spacetime geometry in such regions is well-approximated by a piece of the two-sided BTZ black hole, and the dual CFT state is similarly close to a part of the thermofield-double state. The argument is similar to that of \cite{Maldacena:2001kr} showing the high-temperature 2-sided AdS-Schwarzschild black hole to be dual to the CFT thermofield double, but relies on local analyses of both the CFT path integral and the dominant bulk saddle.  It is possible that the two boundaries shown in figure \ref{fig:BTZRegion} may be connected by parts of the CFT outside the region shown, in which case the thermofield-double entangles two pieces of the same CFT in different parts of space.\footnote{In a two-dimensional CFT, we can act with local conformal symmetries on the state, corresponding to changing the cutoff in the bulk spacetime. To specify the state we therefore must fix the conformal frame, and here we always work in the frame in which the energy density is constant.} Indeed, we will make use of the freedom to entangle regions of the CFT with arbitrary separations to conclude in section \ref{sec:Discussion} that both CA and CV require the complexity of efficient quantum circuits building local thermofield-double-like entanglement of thermal-sized patches to be independent of the separation between patches.

\section{Complexity of formation for AdS wormholes}
\label{sec:GenusBH}

In this section, we compute the complexity of formation of AdS$_3$ wormhole geometries of from both the CV and CA conjectures. The result for spacetimes with $n$ boundaries, and $g$ handles in the causal shadow, takes the form $\Delta C = \alpha c (2-2g-n)$ relative to the complexity of $n$ massless one-sided BTZ black holes, with a coefficient $\alpha$ that we will compute in each case.

\subsection{Volume}

The wormhole geometries we consider are formed as a quotient\footnote{Strictly speaking, the quotient acts not on all of AdS$_3$, but only a subset. This distinction will not be important for our purposes.} of AdS$_3$ under a discrete, finitely generated group of isometries $\Gamma$, and we restrict to those with a time-reflection symmetry, which means that $\Gamma$ commutes with $t\mapsto -t$ \cite{Brill:1995jv,Aminneborg:1997pz,Brill:1998pr}. All the elements of $\Gamma$ fix the $t=0$ slice of AdS$_3$, which we denote $\mathbb{H}^2$ since it has the induced geometry of the hyperbolic plane. The isometries in $\Gamma$ act on $\mathbb{H}^2$ by M\"obius maps with real coefficients, using the half-plane model, so $\Gamma$ forms a Fuchsian group (a discrete subgroup of $SL(2,\mathbb{R})$). The quotient of this $t=0$ slice is the hyperbolic surface $\Sigma = \mathbb{H}^2/\Gamma$, which is the initial data slice of the wormhole geometry. Due to the reflection symmetry, $\Sigma$ has vanishing extrinsic curvature and is thus extremal, and is in fact the only extremal surface anchored to $t=0$ on each boundary. It is therefore this surface whose volume enters the CV conjecture.

To calculate the volume, split the surface up into $n$ exterior pieces $\Sigma_i$ (for $i=1,2,\ldots,n$), and the remaining interior piece $\Sigma_I$:
\begin{equation}
\label{eq:SigmaPart}
\Sigma = \Sigma_I \cup \Sigma_{1} \cup \dots \cup \Sigma_{n}~.
\end{equation}
Each $\Sigma_i$ is the intersection of $\Sigma$ with the domain of outer communication for the $i$th boundary, bounded by the unique geodesic on $\Sigma$ homotopic to the appropriate boundary, the bifurcation surface for the event horizon.  Each $\Sigma_i$ is isometric to the $t=0$ surface of the exterior BTZ solution of some mass $M_i$, or equivalently half the $t=0$ surface for the corresponding two-sided BTZ eternal black hole.  The $t=0$ volume of such $\Sigma_i$ was found in \cite{Chapman:2016hwi} to be independent of $M_i$, and thus to agree with the $M=0$ BTZ black hole.

The remainder $\Sigma_I$ is the interior of the black hole, the ``causal shadow'', which is not causally connected to any exterior region. Its volume is readily computed from the Gauss-Bonnet theorem, since it has constant negative curvature $R^{(2)} = -2/\ell^2$, and is bounded by geodesics, so
\begin{equation}
\label{eq:chiVol}
4\pi \chi  = \int_{\Sigma_I} R^{(2)}\sqrt{|h|}d^2x= -\frac{2}{\ell^2} \text{Volume}(\Sigma_I),
\end{equation}
where $\chi$ is the Euler character of $\Sigma_I$. Using the Brown-Henneaux value $c = \frac{3\ell}{2G}$ of the central charge of the dual CFT, we find the result:
\begin{equation}
\Delta C = \frac{\text{Volume}(\Sigma_I)}{G\ell} = -\frac{4 \pi \chi}{3} c.
\end{equation}
In terms of the genus $g$ and number of boundaries $n$, we have $\chi = 2 -2g-n$. Note in particular that (unless $g=0$ and $n=2$, corresponding to eternal BTZ) $\chi $ is negative, and more negative with increasing genus. Related observations were made recently in \cite{Abt:2017pmf} when discussing subregion complexity.

\subsection{Action}

To compute the action appropriate for the CA conjecture at $t=0$, we must first understand the associated Wheeler-de Witt (WdW) patch $\WdW$.  Recall that the WdW patch anchored at $t=0$ can be defined as the domain of dependence of any AdS-Cauchy surface anchored on the $t=0$ set of the conformal boundary, in particular the surface $\Sigma$ of time symmetry. Since $\Sigma$ is the quotient $\mathbb{H}^2/\Gamma$ of the $t=0$ slice in global AdS$_3$, the Wheeler-de Witt patch $\WdW_\Gamma$ of our wormhole spacetime is the quotient $\WdW_{\rm \!\!AdS_3}/\Gamma$ of the corresponding patch\footnote{Before taking the quotient by $\Gamma$, the region we must excise from AdS$_3$ is the future and past of the limit set of $\Gamma$, a subset of the boundary at $t=0$. This is outside $\WdW_{\rm \!\!AdS_3}$, so is unimportant here.} in global AdS$_3$.

This representation also gives a simple, explicit description of the geometry of $\WdW_\Gamma$. As is well-known, $\WdW_{\rm \!\!AdS_3}$ can be sliced with hyperbolic planes (of varying radii) to admit a metric of FRW form,
\begin{equation}
\label{eq:WdWmetric}
\frac{ds^2}{\ell^2} = - dt^2 + \cos^2t\; d\Omega_{-1}^2,
\end{equation}
where $d \Omega_{-1}^2$ is a metric on the unit-radius $\mathbb{H}^2$, and  $-\frac{\pi}{2}<t<\frac{\pi}{2}$.  As already described, $\Gamma$ acts on the $t=0$ surface by $SL(2,\mathbb{R})$ maps, and in the coordinates of \eqref{eq:WdWmetric}, these isometries extend to act in the same way on all slices of constant $t$.  As a result, the quotient patch $\WdW_\Gamma=\WdW_{\rm \!\!AdS_3}/\Gamma$ is again described by \eqref{eq:WdWmetric}, only replacing $d \Omega_{-1}^2$ by the hyperbolic metric on $\Sigma = \mathbb{H}^2/\Gamma$.

\begin{figure}[t]
\begin{center}
\includegraphics[width=0.4\textwidth]{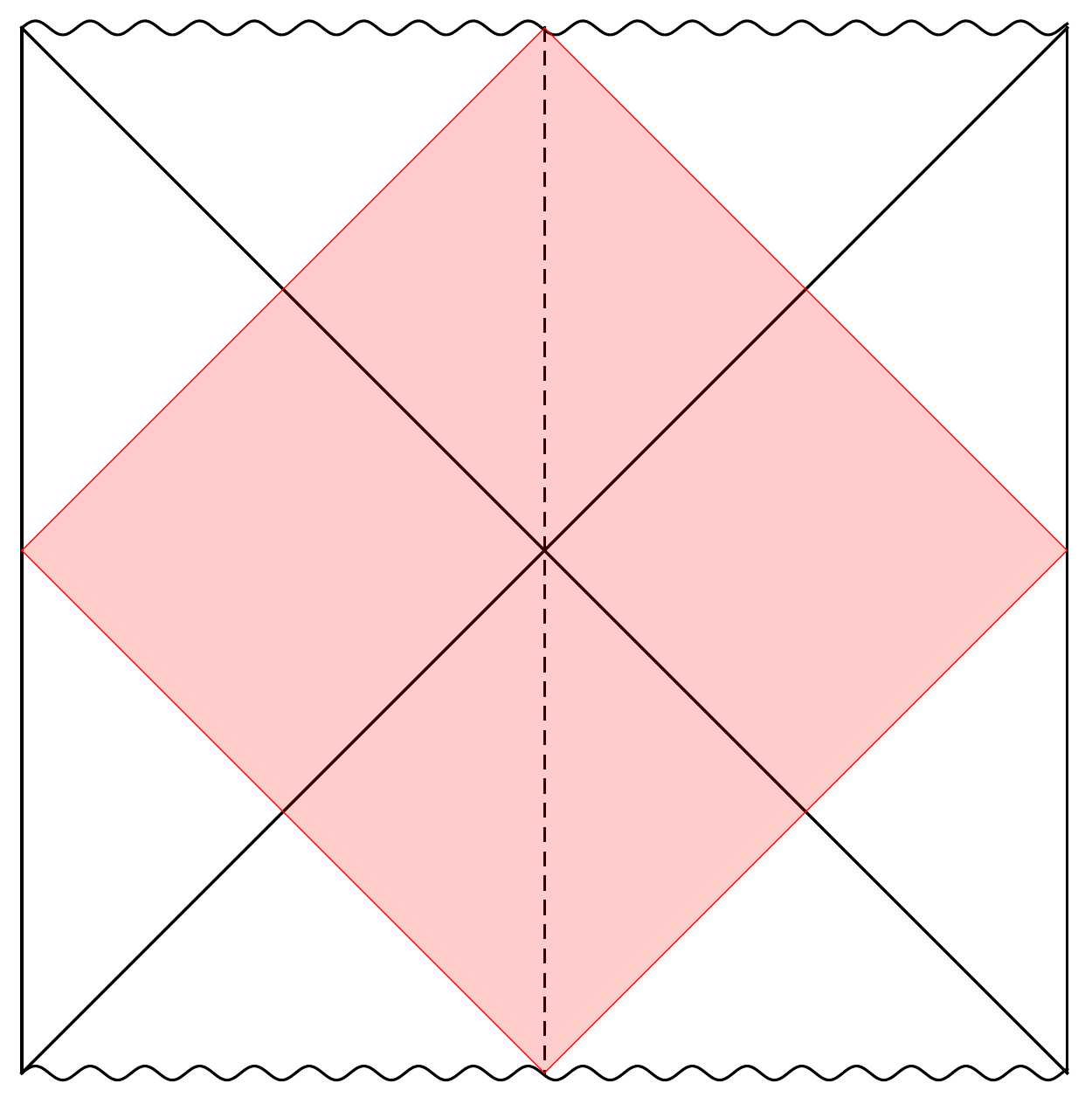}
\end{center}
\caption{The familiar conformal diagram for a 2-sided BTZ wormhole and its $t=0$ WdW patch (shaded).  The dashed vertical line indicates the plane invariant under a right/left reflection symmetry.  The geometry of any $W_i$ is identical to that of this WdW patch to (say) the left of the dashed line.}
\label{fig:HalfBTZ}
\end{figure}

In parallel with our discussion of volumes, we now partition $\WdW_\Gamma$ into non-intersecting pieces $W_i = W_1, \dots W_n$ associated with each boundary and a remaining interior piece $W_I$. Simply by taking the corresponding piece of $\Sigma$ in \eqref{eq:SigmaPart}, times the interval $t\in \left(-\frac{\pi}{2},\frac{\pi}{2}\right)$, so we include the scaled copies of $\Sigma=\mathbb{H}^2/\Gamma$ appearing at each slice of constant $t$. Since $\Sigma_i$ is half of the $t=0$ surface for a 2-sided BTZ wormhole, the geometry on any $W_i$ is precisely that of one half of the WdW patch for this same 2-sided BTZ wormhole, illustrated in figure \ref{fig:HalfBTZ}.

Since the action functionals described in \cite{Lehner:2016vdi} are additive, we may compute the action of $\WdW_\Gamma$ by summing the actions of $W_I, W_1, \dots W_n$.  From the above discussion it also follows that the action of each $W_i$ is precisely half the WdW-patch action for the 2-sided BTZ wormhole, which is independent of the black hole mass \cite{Chapman:2016hwi}.  Noting that the length of such wormholes diverge as $M \rightarrow 0$, and that in this limit the spacetime effectively breaks into two disconnected $M=0$ pieces, it is convenient think of each $W_i$ as having the same action as the WdW patch for a \emph{single} $M=0$ black hole.\footnote{If necessary we may take the above limit as a regulating procedure to define the $M=0$ WdW-patch action.}

The difference $\Delta C$ between the complexity of our wormhole and that of $n$ $M=0$ BTZ black holes is thus determined entirely by the action of $W_I$.  Since each timelike boundary of $W_I$ has the same geometry as the surface indicated by the dashed line in figure \ref{fig:HalfBTZ}, they have vanishing extrinsic curvature, and in any case would be cancelled by matching boundary terms from $W_i$, so do not contribute. To check for contributions to the Gibbons-Hawking boundary term coming from the past and future boundaries $t = \pm \pi/2$, we evaluate it on the regulated spacelike hypersurface at some $t$, which we take to $\pm\pi/2$ at the end. Writing $d\Sigma$ as the volume form on the surface $\Sigma$ with unit negative curvature, so the induced volume form on the slice of constant $t$ is $\ell^2\cos^2 t$ times this, the Gibbons-Hawking term is
\begin{equation}
I_{\rm GH}(W_I) =-\frac{\ell^2}{8\pi G }\lim_{t \to \pm\frac{\pi}{2}}\int_{\Sigma_I} d\Sigma  \;K\cos^2 t,
\end{equation}
where $K$ is the trace of the second fundamental form. The outward-pointing unit normal vector to a constant $t$ surface is $n=\pm\ell^{-1}\frac{\partial}{\partial t}$, and $K$ can be computed as the divergence of this vector field (extended with unit length off the surface):
\begin{equation}
K=\mp \frac{2}{\ell}\tan t.
\end{equation}
This mean curvature tends to infinity, but the intrinsic volume form goes to zero faster, $\cos^2t \tan t \rightarrow 0$ as $t \rightarrow \pm \frac{\pi}{2}$, so these boundary contributions also vanish.

It remains only to compute the bulk contributions of the action of $W_I$ from the Einstein-Hilbert and cosmological terms in \eqref{eq:CA}. Since the spacetime Ricci scalar $R$ and cosmological constant $\Lambda$ are constant, satisfying $R-2\Lambda = -4/\ell^2$, this is proportional to the spacetime volume. In turn, after doing the integral over $t$, this is proportional to the spatial volume of $\Sigma_I$, which we have already computed from the Gauss-Bonnet theorem:
\begin{equation}
I_{\rm EH} (W_I)=\frac{1}{16\pi G}\int d^3x \sqrt{g}(R-2\Lambda) = -\frac{1}{4\pi G \ell}  \left( \int_{-\frac{\pi}{2}}^{\frac{\pi}{2}} dt \cos^2 t \right) \text{Volume}(\Sigma_I).
\end{equation}
Finally, we find the result
\begin{equation}
\Delta C = \frac{1}{\pi} I_{\rm EH} (W_I)= -\frac{\text{Volume}(\Sigma_I)}{8\pi\ell G} = \frac{c}{6}\chi,
\end{equation}
where we again recall that $\chi = 2 -2g-n$ for $n$ boundaries and $g$ handles.

Note that this formula also gives the correct result for pure AdS \cite{Chapman:2016hwi}, $\Delta C = \frac{c}{6}$. Our formula for the CV conjecture gives $\Delta C =-\frac{4\pi}{3}c$ for pure AdS, also agreeing with the result of \cite{Chapman:2016hwi}.

\section{Nonlocality of complexity}
\label{sec:Discussion}

We observed above that the $t=0$ volume and action of $n$-boundary geon spacetimes depend only on $n$ and the genus $g$, and are independent of the black hole temperatures and internal moduli.  We found the $n$-boundary wormhole with Euler characteristic $\chi$ to have complexity
$\Delta C = \alpha c \chi$ relative to $n$ $M=0$ BTZ black holes, in terms of the central charge $c$ of the dual field theory, with coefficients $\alpha_V = -4\pi /3$, $\alpha_A = 1/6$ for CV and CA conjectures respectively.  This suggests that any CFT gate set defining either complexity necessarily contains bi-local gates acting at points with arbitrary separation, as we will now argue.

In particular, we now show that the complexity of an efficient quantum circuit building local thermofield-double-like entanglement of thermal-sized patches cannot depend on the separation of the patches so entangled. To show this, we will construct a class of pure states on a single CFT in which entangled patches can be at arbitrary separation, by altering the moduli at fixed genus, and hence fixed complexity according to either CA or CV conjectures.

\begin{figure}[t]
\begin{center}
\includegraphics[width=0.5\textwidth]{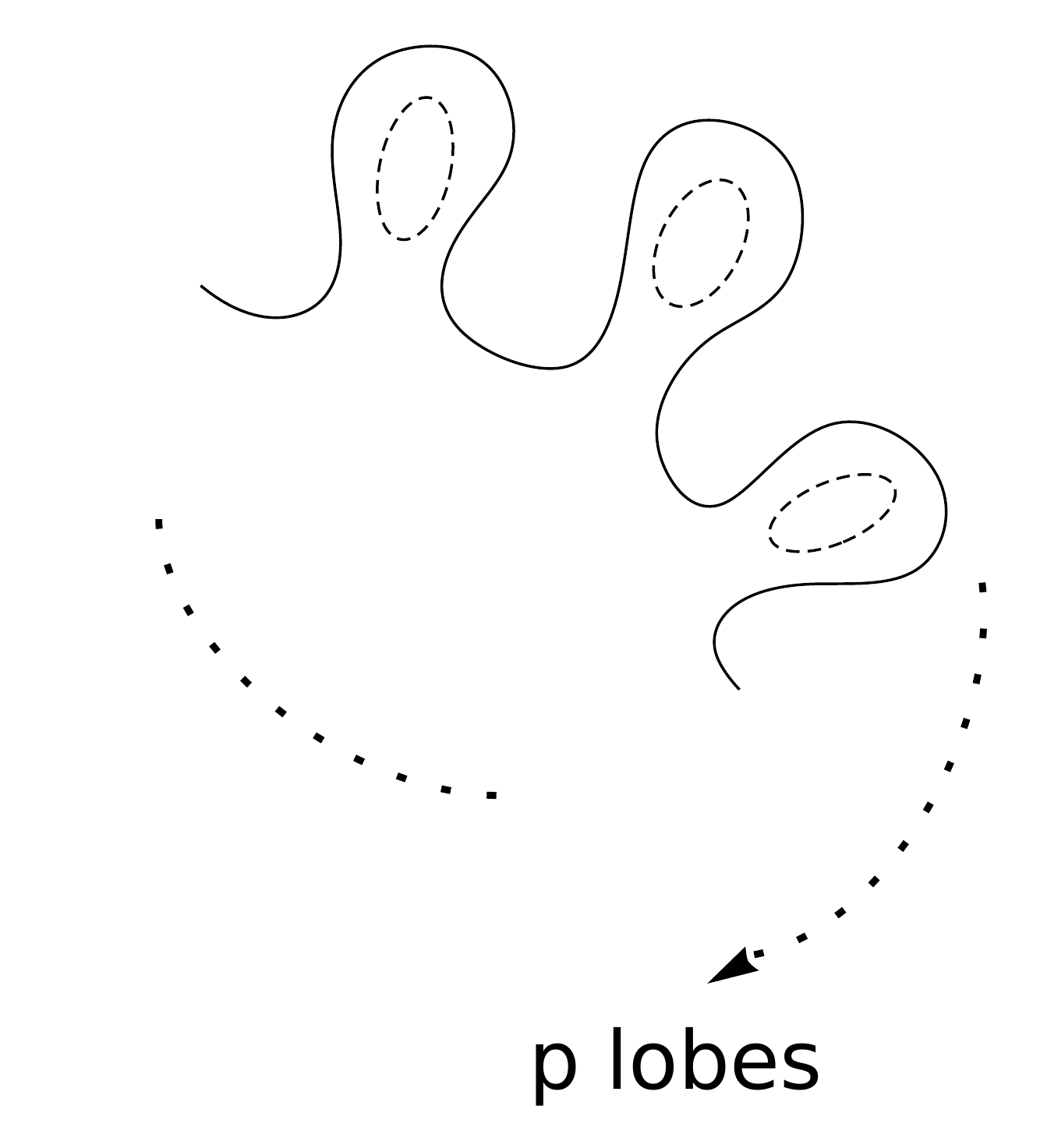}
\end{center}
\caption{A $t=0$ surface with $p+1$ boundaries (one solid line and $p$ dashed lines).}
\label{fig:lobes}
\end{figure}

 To make this explicit, we first construct a particularly convenient class of 1-boundary ($n=1$) wormholes in which the $t=0$ causal shadow is built from $2p-2$ triangular islands of the form shown in figure \ref{fig:islands} (left) for some even integer $p> 1$. Note that $p-2$ of these triangles can be connected in a tree structure to form a region from which $p$ tendrils extend.  On each of the remaining $p$ triangles, we connect two tendrils together in a loop to make $p$ copies of the the region shown in figure \ref{fig:islands} (right).  We then attach one copy of figure \ref{fig:islands} (right) to each of the $p$ tendrils extending from the tree built in the first step.  The result at this stage is a spacetime with $p+1$ boundaries and a genus zero $t=0$ surface of the form shown in figure \ref{fig:lobes}.

The final step in our construction is to use the fact that $p$ is even to group the $p$ boundaries drawn in figure \ref{fig:lobes} as dashed lines into $p/2$ pairs, and to identify the surface in figure \ref{fig:lobes} along such pairs of dashed lines.  The resulting diagram then defines a $t=0$ surface with only a single remaining boundary (solid line) and genus $g=p/2$.  The genus, and thus both volume and action, depend only on $p$ and not on the particular pairing chosen.  In particular, if we number the dashed boundaries sequentially with elements of ${\mathbb Z}_p$, identifying adjacent dashed boundaries (say number $2k$ and $2k+1$) gives precisely the same result as identifying diametrically opposed dashed boundaries ($k$ and $k+p/2$), even though the typical distance spanned by thermofield-double-like entanglements in the latter case is much longer.  Since the states are otherwise identical, we conclude that the complexity of an efficient quantum circuit building local thermofield-double-like entanglement of thermal-sized patches is indeed independent of the patches' separation.

Finally, we turn to interpreting the coefficients $\alpha_V = -4\pi/3$, $\alpha_A = 1/6$.  Here is it useful to recall from \cite{Chapman:2016hwi} that under either CV or CA the BTZ complexities are also independent of temperature.  So we may think of our $\Delta C$ as being defined relative to a collection of (half-) BTZ-wormholes of the same temperatures as the black holes at the ends of our wormhole.  Since most of the dual quantum state is locally a thermofield double, and having established above that the complexity of thermofield-double-like regions does not depend on the separation between the regions they entangle, we see that $\Delta C$ directly measures the complexity associated with the $q$ regions of our state that are {\it not} locally of thermofield-double form.  In particular, it measures the complexity of these $q$ regions relative to $q/2$ thermofield-double regions of equivalent size.

In fact, we know from \cite{Marolf:2015vma} that when the triangles discussed in section \ref{sec:overview} are well-separated, each of these $q$ regions has roughly thermal size and we have $q=3m$ where $m$ is the (integer) number of such triangles. Furthermore, such regions are entangled in $m$ groups of $3$ in essentially the same way as 3 regions of the CFT vacuum spaced equally around the circle on which the CFT lives.  We refer to the associated 3-party state as a triangular state below.

It is therefore useful to write our results in terms of $m$.  In the above limit, the volume of the causal shadow region is clearly linear in $m$.  From \eqref{eq:chiVol} and any simple example one thus finds $m= -2\chi$. For example, the the $n=2$ wormhole with $g=0$ requires $m=0$ has no causal shadow and so uses $m=0$ triangles, while the $n=3$ wormhole with $g=0$ has $\chi = -1$ and  requires $m=2$ triangles.  It follows that $\Delta C_V = 2\pi m c/3$ while $\Delta C_A = - mc/12$.  The negative value of $\Delta C_A/m$ then indicates that the triangular state is less complex than $3/2$ locally thermofield-double states.  In particular, while one might have thought that all $n$-party entanglement would be built by combining 2-party entanglements in complicated ways, at least in the CA version of complexity we find that two-party entanglement (of the thermofield-double sort) cannot be viewed as more fundamental than 3-party entanglement. A priori, this is a rather different non-locality than that argued  in \cite{Hashimoto:2017fga} to be needed to match holographic complexity, but it would be interesting to more carefully investigate any possible connections.

\section{A bound on genus?}\label{genusBound}

A surprising feature is the difference in sign between the CA and CV conjectures. A similar result was found for the BTZ black hole, where the CA conjecture leads to a negative complexity relative to that of the vacuum \cite{Chapman:2016hwi}, and in \cite{Reynolds:2017jfs} which studied compactifications of $d=3$ CFTs on $S^1$.  In all cases the difference in sign comes from the fact that increasing the spacetime volume clearly increases CV, while the negative cosmological constant means that this tends to decrease the volume part of the action for CA.  Note also that the contribution to CA from singularities vanishes in these cases.

The negative for answer was not particularly disturbing, as it can be attributed to an additive ambiguity in the definition of complexity which is associated  to the choice of reference state. In  the present case the sign difference appears much more significant, as it multiplies the genus $g$ of the solution.  Thus no matter what reference state is chosen, i.e. no matter what additive constant we include in the complexity, the CA complexity appears to become negative at sufficiently high genus. In this section, we will discuss the possibility that this apparent contradiction should be evaded, and positivity of the complexity guaranteed, by an upper bound on the genus.

We begin with the qualitative observation that the complexity of a state will be sensitive to the UV structure of the theory, and will naturally contain a (large and positive) UV-sensitive additive constant. which should be proportional to the volume of space in units of the UV cutoff.  This is easiest to see in a lattice regularization, where -- if we define the complexity relative to a product state -- the complexity should be proportional to the number of lattice sites. This UV divergent piece should in addition scale with the number of local degrees of freedom per point in space, hence be proportional to the central charge $c$. This is qualitative result that would be obtained by applying a na\"ive cutoff to the bulk conjectures, though it is unclear what ``counterterms'' depending on the intrinsic geometry of the cutoff should be added. The result is a complexity, counting the number of gates to prepare the CFT from some reference state (that does \emph{not} resemble the CFT vacuum in the UV), of the form
\begin{equation}
C \approx k c \frac{L}{\epsilon} +\Delta C
\end{equation}
for some order one constant $k$, with $L$ the spatial size of the system and $\epsilon$ the cutoff (specializing to $(1+1)$-dimensional systems). There may be other constant terms, as would arise for example if we chose to compute $\Delta C$ relative to the vacuum instead of a thermofield double state, but we ignore these as they are negligible in the limit $\epsilon\ll L$. Unlike $\Delta C$, this quantity should be positive, providing a lower bound on $\Delta C$, proportional to $-c \frac{L}{\epsilon}$.

From the CA results, if we take the genus to be of order $\frac{L}{\epsilon}$, it appears we can violate this bound. But it may not be sensible to talk about the state dual to the geon when the genus is this large. Indeed, in the high-temperature limit we have focussed on, the state is described by order $g$ regions entangled in thermofield-double type states. But for this description to make sense, the regions must be large compared to the thermal scale, which implies a lower bound on the effective temperature, $TL\gtrsim g$, to fit $n$ such regions into the boundary. The result is that to violate the bound requires a temperature of at least the UV scale, $T\gtrsim \epsilon^{-1}$, so the state cannot be faithfully represented with that value of the cutoff.

Having made this argument in the hot limit, it may seem problematic that the geon geometries exist far away from the limit, at any temperature for given genus, by altering the moduli. However, the dual description of the state undergoes first-order phase transitions, generalizing the Hawking-Page transition, so the geon may not be the dominant saddle-point in any semiclassical path integral. On this basis, one might propose that for given genus, there is a lower bound on the energy required for a geon of that genus to dominate the path integral. Furthermore, this minimum energy scales linearly with the genus at large $g$, so (in CFT units) $E\geq A c g$ for some constant $A$.

We can explicitly derive such a bound in a particular symmetric example, with a single boundary, genus $g$, $\mathbb{Z}_n$ symmetry with $n=2g+1$, and a single modulus that we can associate with the energy of the state. It is easiest to describe the $t=0$ slice $\Sigma$ of the geometry as an $n$-fold cover of a disc, branched at two points. After going around either of these points anticlockwise, move ``up'' to the next sheet, so going round the whole boundary, which involves passing round both points, takes you up two sheets. Since $n$ is odd, you will only return to your starting point after going round $n$ times, so the cover has a single boundary circle. This describes the topology; for a smooth metric on the cover, the two branch points on the disc must have conical defects of opening angle $2\pi/n$, and there is a one-parameter family of complete hyperbolic surfaces with this topology and such conical defects (just as there is a one-parameter configuration space of a pair of points on the hyperbolic plane, up to symmetries\footnote{To prepare this state in the CFT, one way would be to consider a $\mathbb{Z}_n$ orbifold theory, and consider the state prepared by the insertion of two twist operators, which have precisely this configuration space. The original CFT Hilbert space is equivalent to the appropriate twisted sector of the orbifold theory.}). There is therefore a single real parameter, which we will identify with the length of the horizon, specifying $\Sigma$, and hence the geon spacetime. In the first nontrivial case $n=3$, this geometry can be thought of as a $\mathbb{Z}_3$ symmetric torus (with modular parameter $\tau=e^{2\pi i/3}$, roughly speaking) hidden behind an event horizon. In the hot limit, the dual CFT state is described by $2n$ regions of equal size, each in a local thermofield double with the region diametrically opposite (though oriented oppositely, so the points entangled with one another still vary discontinuously, unlike for the $\mathbb{RP}^2$ geon). This geometry is a close relative of a $\mathbb{Z}_n$ symmetric wormhole with $n$ boundaries and zero genus; one simply takes a different cover, where you go up a sheet after going anticlockwise around one of the points, and clockwise around the other.

For some value of the horizon radius, which can be computed numerically, there is a phase transition to a pure AdS geometry, since there is an alternative saddle-point to the Euclidean path integral with the same boundary conditions. The boundary conditions are that the geometry is asymptotic to an appropriate genus $2g$ surface, obtained by gluing $\Sigma$ to a reflected version of itself along its boundary. Assuming the dominant geometry respects the $\mathbb{Z}_n$ symmetry, we may quotient by this symmetry, so this is equivalent to finding a three-manifold obeying the Euclidean equations of motion, with boundary Riemann sphere, and two lines of $\frac{2\pi}{n}$ conical defects joining four marked points on the boundary in pairs (two copies of the branch points of the cover). The (real) cross-ratio $x$ of these four points is one way to parameterize the modulus of the geometry. Cyclically permuting the four boundary points maps between a solution with cross-ratio $x$, and one with cross-ratio $1-x$, which implies that the phase transition happens at $x=\frac{1}{2}$ (analogously to the Hawking-Page transition occuring at $\tau=i$, the fixed point of the modular transformation $\tau\mapsto-\frac{1}{\tau}$). Having found the solution with $x=\frac{1}{2}$, the horizon length is just $n$ times the length of a geodesic encircling the two conical defects. In the $n\to\infty$ limit, the conical defects become cusps, receding to infinite distance, but the solution with a pair of cusps joining the boundary points still exists, and the geodesic going around the cusps approaches some finite length $L$. The horizon length in the $n\to\infty$ limit therefore asymptotically approaches $n L$. It is possible to numerically compute $L\approx 8.02$, but for us the only salient point is the scaling proportional to $n$. To summarize, in this symmetric family of states, the dominant geometry is the geon with genus $g$ interior only when the horizon size in AdS units (or energy in units of $c$, or temperature for fixed boundary size) is larger than an order one constant times $g$.

The proposed result is somewhat similar to a Bekenstein bound from entropy considerations. If there is a very high genus single-exterior black hole, the moduli of the internal geometry (or the states of matter living on this geometry) will encode many states, and this could exceed the number of possible states from the Bekenstein-Hawking entropy; this was worked out explicitly in \cite{Maloney:2015ina}. This means that these geometries cannot possibly describe distinct semiclassical states, bounding $g\lesssim S$. However, the conjectured bound here is far stronger, since though it scales the same way with energy, the entropy is proportional to $c$, rather than the order one coefficient suggested here.

\section*{Acknowledgements}
It is a pleasure to thank Adam Brown, Rob Myers, Massimo Porrati, Dan Roberts and Leonard Susskind for useful discussions. ZF, AM, DM, and HM were supported by the Simons Foundation. ZF and DM were also supported in part by funds from the University of California. AM was also supported by the National Science and Engineering Council of Canada. ZW was funded by the Pilot Scheme of Talent Training in Basic Sciences (Boling Class of Physics, Nankai University), Ministry of Education of the People's Republic of China. ZW thanks the UCSB Physics department for its hospitality during the bulk of this work.

\bibliographystyle{jhep}
	\cleardoublepage
\phantomsection
\renewcommand*{\bibname}{References}

\bibliography{GC}

\end{document}